\documentclass[conference]{ifacconf}

\usepackage{graphicx}
\usepackage{url}
\usepackage{textcomp}
\usepackage{mathtools}
\usepackage{tikz}
\usepackage[americanresistors,americaninductors]{circuitikz}
\usepackage{textcomp}
\usepackage{tikz}
\usepackage[tight]{subfigure}
\usetikzlibrary{dsp,shapes,arrows, calc, chains, positioning, quotes, shadows}
\tikzstyle{block} = [draw, rectangle, 
    minimum height=3em, minimum width=4em,align=center]
\tikzstyle{hblock} = [draw, rectangle, 
    minimum height=15em, minimum width=3em,align=center]
\tikzstyle{sum} = [draw, circle, node distance=1cm]
\tikzstyle{input} = [coordinate]
\tikzstyle{output} = [coordinate]
\tikzstyle{pinstyle} = [pin edge={to-,thin,black}]

\usepackage{amsmath}
\usepackage{natbib}
\usepackage{verbatim}
\usepackage{listings}
\usepackage{amssymb}

{\theorembodyfont{\slshape}\newtheorem{theorem}{Theorem}[section]}
{\theorembodyfont{\slshape}\newtheorem{proposition}[theorem]{Proposition}} 
{\theorembodyfont{\slshape}}
{\theorembodyfont{\slshape}}           %
{\theorembodyfont{\slshape}}
{\theorembodyfont{\slshape}}  %
{\theorembodyfont{\upshape}\newtheorem{definition}[theorem]{Definition}}
{\theorembodyfont{\upshape}}
{\theorembodyfont{\upshape}\newtheorem{remark}[theorem]{Remark}}
{\theorembodyfont{\upshape}}
\let\cite\citep

\begin{document}
\begin{frontmatter}
\title{A recursive estimation approach to distributed identification of large-scale multi-input-single-output FIR systems\thanksref{footnoteinfo}}

\thanks[footnoteinfo]{This work was supported by the European Research Council (ERC), under the grant SYSDYNET (grant agreement \text{\textnumero} 694504).}

\author[First]{Tom R.V. Steentjes,}
\author[First]{Mircea Lazar,}
\author[First]{Paul M.J. Van den Hof}

\address[First]{Department of Electrical Engineering, Eindhoven University of Technology, The Netherlands (e-mails:  t.r.v.steentjes@tue.nl, m.lazar@tue.nl, p.m.j.vandenhof@tue.nl)}

\begin{abstract}
The problem of identifying single modules in multiple-input-single-output (MISO) systems is considered. A novel approach to distributed identification of MISO finite impulse response systems is presented. The distributed  identification is discerned by the local estimation of local parameters, which correspond to a module in the MISO system. The local estimators are derived from the standard recursive least squares estimator and require limited information exchange. By Lyapunov's second method, sufficient conditions are derived for asymptotic convergence of the estimators to the true parameters in the absence of disturbances, which lead to asymptotic unbiasedness in the presence of additive output disturbances.
\end{abstract}
\end{frontmatter}
\section{Introduction}
Prediction-error identification methods provide a powerful tool for obtaining consistent system parameter estimates \cite{ljung1999}. However, when dealing with large scale interconnected systems, such as the ones arising from biology or power grids, the identification problem becomes more challenging. Given a network of linear dynamical systems, various prediction error methods are readily operational for identifying these systems \cite{rao1984}, \cite{vandenhofetal2013}.


The identification problem of such large-scale systems can typically be separated into multiple-input-single-output (MISO) identification problems \cite{rao1984}, \cite{vandenhofetal2013}. More precisely, identification of a large-scale system can be performed via the identification of MISO building blocks, on the basis of measurements of multiple inputs and one, possibly disturbed, output. Figure \ref{fig:m7scheme} shows such a MISO building block.

Although existing prediction error methods for dynamical networks can consistently identify local modules (single-input-single-output (SISO) systems), they require the output signal and all input signals for a MISO identification problem to be available centrally for global parameter estimation. Central data collection and computation of the module estimates may not always be desirable due to computational constraints or desired flexibility. A further decomposition of the MISO identification problem into SISO identification problems to reduce computational complexity was also suggested in \cite{rao1984}. Therein, it was proposed to perform a decomposition of the parameter estimation via a Gauss-Seidel like algorithm, but a proof of convergence is absent.

Distributed estimation has caught a vast amount of attention in the literature. Existing approaches can be divided into two distinct classes. The first class consists of consensus based methods, discerned by collaborative estimation of a global (common) parameter vector that is performed via a number of interconnected estimators \cite{mateos2012}, \cite{papusha2014}. The second class is also enabled by collaborative estimation via interconnected estimators. Therein each estimator is, however, concerned with the estimation of a local parameter vector. We refer to the results derived for parameter estimation in static large-scale systems \cite{marelli2015}, distributed state estimation via moving-horizon methods \cite{farina2010} and distributed identification via ADMM \cite{hansson2014}.

In this work, we develop a distributed solution for the MISO prediction error identification problem \cite{vandenhofetal2013}. Due to the simplifying property of yielding output predictors that are linear in the parameters, finite impulse response (FIR) model structures serve as a basis for the developed distributed identification method. The distributed identification scheme is composed of local recursive estimators that are coupled with local SISO modules. Intercommunication of the local estimators is accomplished through the transmission of scalar signals between recursions via a mutual fusion center. 


\section{Preliminaries and problem formulation}
The sets of non-negative integers and non-negative reals are denoted by $\mathbb{N}$ and $\mathbb{R}_{\geq 0}$, respectively. Given $a\in \mathbb{N}$, $b\in \mathbb{N}$ such that $a<b$, we denote $\mathbb{N}_{[a:b]}:=\left\{a,a+1,\dots,b-1,b\right\}$. Let $I_n\in\mathbb{R}^{n\times n}$ denote the identity matrix. A function $\alpha:\mathbb{R}_{\geq 0}\rightarrow \mathbb{R}_{\geq 0}$ is said to belong to class $\mathcal{K}$ ($\alpha\in \mathcal{K}$), if it is continuous, strictly increasing and $\alpha(0)=0$. It is said to belong to class $\mathcal{K}_\infty$ ($\alpha\in \mathcal{K}_\infty$), if additionally $\alpha(r)\rightarrow \infty$ as $r\rightarrow \infty$. For an $x\in \mathbb{R}^n$, let $\|x\|_2$, or simply $\|x\|$, denote the $2$-norm of $x$.

\subsection{Concepts from Lyapunov theory}
Consider the discrete-time, time-varying system
\begin{align} \label{eq:fdt}
x(t+1)=f(x(t),t),\quad x_0:=x(t_0),\ t_0\in\mathbb{N},
\end{align}
with $f:\mathbb{R}^n\times \mathbb{N}\rightarrow\mathbb{R}^n$, $f(0,\cdot)=0$. Let the solution of \eqref{eq:fdt} initialized in $x_0\in\mathbb{R}^n$ at time $t_0\in\mathbb{N}$ be denoted by $s(t,t_0,x_0)$.
\begin{definition}
The origin equilibrium of \eqref{eq:fdt} is called stable if for each $\varepsilon>0$ and each $t_0\in\mathbb{N}$, there exists $\delta=\delta(\varepsilon,t_0)$ so that
\begin{align*}
\|x_0\|<\delta\ \Rightarrow\ \|s(t,t_0,x_0)\|<\varepsilon,\quad \forall t\geq t_0.
\end{align*}
\end{definition}
\begin{definition}
The origin equilibrium of \eqref{eq:fdt} is called attractive if there is a $\delta>0$ such that
\begin{align}
&\text{For each } \varepsilon>0 \text{ there exists } T=T(\varepsilon,t_0) \text{ such that }\nonumber\\
&\|x_0\|<\delta\ \Rightarrow\ \|s(t,t_0,x_0)\|<\varepsilon,\quad \forall t\geq t_0+T. \label{def:att}
\end{align}
\end{definition}
By the definition of a function limit at infinity, \eqref{def:att} is equivalent with:
\begin{align*}
\|x_0\|<\delta\ \Rightarrow\ \|s(t,t_0,x_0)\|\rightarrow 0 \text{ as } t\rightarrow \infty.
\end{align*}
\begin{definition}
The origin equilibrium of \eqref{eq:fdt} is called globally attractive if
\begin{align*}
x_0\in\mathbb{R}^n\  \Rightarrow \|s(t,t_0,x_0)\|\rightarrow 0 \text{ as } t\rightarrow \infty.
\end{align*}
\end{definition}
\begin{definition}
The origin equilibrium of \eqref{eq:fdt} is called asymptotically stable if it is stable and attractive.
\end{definition}
\begin{definition}
The origin equilibrium of \eqref{eq:fdt} is called globally asymptotically stable if it is stable and globally attractive.
\end{definition}
\begin{theorem} \label{thm:stab}
The origin is a stable equilibrium of \eqref{eq:fdt} if there is a function $W:\mathbb{R}^n\times\mathbb{N}\rightarrow\mathbb{R}$, so that
\begin{align}
k_1(\|\xi\|)\leq W(\xi,\tau)\leq k_2(\|\xi\|,\tau),\quad &\forall (\xi,\tau)\in\mathbb{R}^n\times \mathbb{N},\label{cond:1a}\\
\Delta W(\xi,\tau)\leq 0,\quad &\forall (\xi,\tau)\in\mathbb{R}^n\times \mathbb{N},\label{cond:1b}
\end{align}
with $k_1\in\mathcal{K}_\infty$, $k_2(\cdot,\tau)\in\mathcal{K}_\infty$ for each $\tau\in\mathbb{N}$ and $\Delta W(\xi,\tau):=W(f(\xi,\tau),\tau+1)-W(\xi,\tau)$.
\begin{pf}
See Appendix A.
\end{pf}
\end{theorem}
\begin{theorem} \label{thm:lyap}
The origin is a globally asymptotically stable equilibrium of \eqref{eq:fdt} if there is a function $W:\mathbb{R}^n\times\mathbb{N}\rightarrow\mathbb{R}$ so that
\begin{align}
k_1(\|\xi\|)\leq W(\xi,\tau)\leq k_2(\|\xi\|,\tau),\quad &\forall (\xi,\tau)\in \mathbb{R}^n\times \mathbb{N},\label{cond:2a}\\
\Delta W(\xi,\tau)\leq -k_3(\|\xi\|),\quad &\forall (\xi,\tau)\in \mathbb{R}^n\times\mathbb{N},\label{cond:2b}
\end{align}
with $k_1\in\mathcal{K}_\infty$, $k_2(\cdot,\tau)\in\mathcal{K}_\infty$ for each $\tau\in\mathbb{N}$ and $k_3:\mathbb{R}_{\geq 0}\rightarrow \mathbb{R}_{\geq 0}$ a continuous and positive definite function.
\begin{pf}
See Appendix B.
\end{pf}
\end{theorem}
Note the absence of a uniform upperbound on $W$ in Theorem \ref{thm:lyap}. This avoids the need for a uniform lower bound on, or termination of gain/covariance matrix recursions as in \cite{mendel73}, \cite{udinktencate79}, for proving convergence of the recursive estimation scheme in Section~\ref{sec:con}.
\begin{definition}
A function $W:\mathbb{R}^n\times\mathbb{N}\rightarrow\mathbb{R}$ that satisfies \eqref{cond:2a} and \eqref{cond:2b} is called a Lyapunov function for \eqref{eq:fdt}.
\end{definition}

\subsection{Prediction error set-up and least squares estimator} \label{sec:prederr}
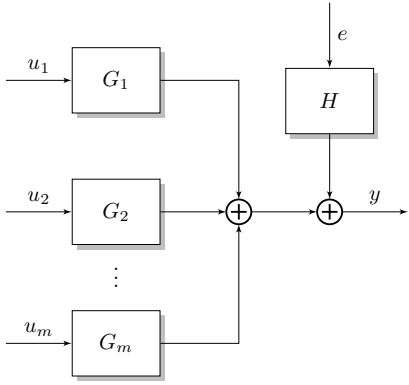
\begin{figure} [!t]
\centering
\resizebox{2.2in}{!}{
\begin{tikzpicture}[auto, node distance=3em,>=latex']
    \node(input)(u2) {};
    \node[input, above= of u2](u1) {};
    \node[input, below= of u2](u3) {};
    
    \node [block,right=of u2, fill = white, drop shadow](G2){$G_2$};
    \node [block,above=of G2, fill = white, drop shadow](G1){$G_1$};
    \node [block,below=of G2, fill = white, drop shadow](G3){$G_m$};
    \node[input, left= of G1](u1) {};
    \node[input, left= of G3](u3) {};    
    
    \node[dspadder,right= of G2](sum1){};
    
    \node[dspadder,right=of sum1](sum2){};
    \node[block,above= of sum2, fill = white, drop shadow](H1){$H$};
    \node[input,above=of H1](e1){};
    \node[input,right= of sum2](y1){};

    \draw[->] (u2)--node{$u_2$} (G2);
    \draw[->] (u1)--node{$u_1$} (G1);
    \draw[->] (u3)--node{$u_m$} (G3);
    
    \draw[->] (G1)-| node[pos=0.95]{$ $} (sum1);
    \draw[->] (G3)-| node[pos=0.95]{$ $} (sum1);
    \draw[->] (G2)-- node[pos=0.85]{$ $} (sum1);
    
    \draw[->] (e1)--node{$e$} (H1);
    \draw[->] (H1)-- node[pos=0.85]{$ $} (sum2);
    \draw[->] (sum1)-- node[pos=0.85]{$ $}(sum2);
    \draw[->] (sum2)-- node{$y$}(y1);
    \node at ($(G2)!.45!(G3)$) {\vdots};
\end{tikzpicture}}
\caption{MISO system interconnection with $m\in\mathbb{N}$ subsystems.}
\label{fig:m7scheme}
\end{figure}

Consider a MISO system interconnection with $m$ inputs and a measured output that is corrupted by a noise signal $v(t)$, described by
\begin{align} \label{eq:sysmiso}
y(t)=\sum_{i=1}^m G_i(q)u_i(t)+v(t),
\end{align}
where $v(t)=H(q)e(t)$, with $e(t)$ zero-mean white noise with standard deviation $\sigma$ for all $t\in\mathbb{N}$. System \eqref{eq:sysmiso} can, for example, appear as a building block in dynamic networks analyzed in \cite{vandenhofetal2013}, cf. \cite[Section~2.1]{vandenhofetal2013}. An example of such a MISO system with $m$ inputs is shown in Figure \ref{fig:m7scheme}.

A prediction error identification problem for identifying $G_i(q)$, $i\in\mathbb{N}_{[1:m]}$,  is based on \cite{vandenhofetal2013}: an output prediction
\begin{align*}
\hat{y}(t,\theta)&:=\hat{H}^{-1}(q,\theta)\left(\sum_{i=1}^m \hat{G}_i(q,\theta)u_i(t)\right)\\
&\qquad+(1-\hat{H}^{-1}(q,\theta))y(t),
\end{align*}
and the prediction error, defined by
\begin{align*}
\varepsilon(t,\theta):=y(t)-\hat{y}(t,\theta),
\end{align*}
where $\hat{G}_i(q,\theta)$ is a model of transfer function $G_i(q)$, $i\in\mathbb{N}_{[1:m]}$, depending on a to-be-estimated parameter vector $\theta\in\mathbb{R}^n$.

Due to the possible independent parametrization and advantageous property of yielding an output prediction $\hat{y}(t,\theta)$ that is linear in the parameters $\theta$, an FIR model structure will be employed in the sequel. Utilizing an FIR model structure, the noise model $\hat{H}(q,\theta) = 1$ is fixed and the plant models are
\begin{align} \label{eq:pm}
\hat{G}_i(q,\theta) = B_i(q,\theta)
\end{align}
for all $i\in\mathbb{N}_{[1:m]}$, where 
\begin{align*}
B_i(q,\theta)&:=b_0^i+b_1^iq^{-1}+\dots +b_{n_i-1}^iq^{-n_i+1},
\end{align*}
with $n_i\in\mathbb{N}$ the number of parameters in the polynomial $B_i(q,\theta)$, $i\in\mathbb{N}_{[1:m]}$
Consequently, the output prediction is
\begin{align} \label{eq:outpred}
\hat{y}(t,\theta)=\sum_{i=1}^mB_i(q,\theta)u_i(t)=\sum_{i=1}^m \varphi_i^\top(t)\theta_i,
\end{align}
with $\theta_i:=\operatorname{col}(b_0^i,\dots, b_{n_i-1}^i)$ and
\begin{align*}
\varphi_i^\top(t)&:=\begin{pmatrix}
u_i(t) & \cdots & u_i(t-n_i+1)
\end{pmatrix},
\end{align*}
such that the total parameter vector is $\theta:=\operatorname{col}(\theta_1,\dots,\theta_m)$ and $\varphi(t):=\operatorname{col}(\varphi_1(t),\dots,\varphi_m(t))$.

Given $N\in\mathbb{N}$ available data samples of $y$ and $\varphi_i$, $i\in\mathbb{N}_{[1:m]}$, let the identification criterion be $\min_\theta J_{\text{LS}}(\theta)$, where $J_{\text{LS}}:\mathbb{R}^n\rightarrow\mathbb{R}$ is defined by
\begin{align*}
J_{\text{LS}}(\theta):=\frac{1}{N}\sum_{t=0}^{N-1} \varepsilon^2(t,\theta)=\frac{1}{N}\|\mathbf{y}-\Phi\theta\|^2_2,
\end{align*}
with vector $\mathbf{y}^\top:=(y(0),\dots,y(N-1))$ and matrix $\Phi^\top:=(\varphi(0),\dots,\varphi(N-1))$. The optimal parameter estimate is explicitly known to  be \cite{kay1993}
\begin{align} \label{eq:LSE}
\hat{\theta}:=\operatorname{arg}\min_\theta J_{\text{LS}}(\theta)=(\Phi^\top\Phi)^{-1}\Phi^\top \mathbf{y}.
\end{align}
The parameter estimate $\hat{\theta}$ is referred to as the \textit{least squares estimator} (LSE). The covariance matrix of the LSE is $\Sigma=\sigma^2(\Phi^\top \Phi)^{-1}$ \cite{kay1993}.

\subsection{Recursive least squares}

In practice, computing the LSE can be undesirable when all the data $\mathbf{y}$ and $\Phi$ are not available at once or when \eqref{eq:LSE} is computationally intractable, for example. Instead, one can use a recursive LSE \cite{kay1993}, which updates the LSE each time new data  is available.

Let $\hat{\theta}(k)$ denote the LSE of $\theta$ based on $k+1$ data samples $\mathbf{y}(k)=(y(0)\ \cdots\ y(k))^\top$ and $\Phi(k)=(\varphi(0)\ \cdots \varphi(k))^\top$. The recursive LSE reads as follows \cite{kay1993}. First, compute the ``batch'' estimator $\hat{\theta}(k)$ for $k\in\mathbb{N}$:
\begin{align}
\hat{\theta}(k)&=(\Phi(k)^\top \Phi(k))^{-1}\Phi(k)^\top \mathbf{y}(k),\label{eq:batchls}\\
\Sigma(k)&=\sigma^2(\Phi(k)^\top \Phi(k))^{-1}.\nonumber
\end{align}
When new data is available, update the estimator according to
\begin{align*}
\hat{\theta}(k+1)&=\hat{\theta}(k)\\
&+\alpha(k)\Sigma(k)\varphi(k+1)(y(k+1)-\varphi^\top(k+1)\hat{\theta}(k)),\\
\alpha(k)&:=\frac{1}{\sigma^2+\varphi^\top(k+1)\Sigma(k)\varphi(k+1)}.
\end{align*}
The covariance matrix of the updated LSE is
\begin{align}
\Sigma(k+1)=(I-\alpha(k)\Sigma(k)\varphi(k+1)\varphi^\top(k+1))\Sigma(k). \label{eq:RLSs}
\end{align}
\begin{remark}
The recursive LSE and covariance matrix can be written in a more compact form, using the prediction error definition and the matrix inversion lemma, as
\begin{align}
\hat{\theta}(k+1)&=\hat{\theta}(k)+\alpha(k)\Sigma(k)\varphi(k+1)\varepsilon(k+1,\hat{\theta}(k)),\hspace{-.2em}\label{eq:RLSt}\\
\hspace{-.5em}\Sigma^{-1}(k+1)&=\Sigma^{-1}(k)+\frac{1}{\sigma^2}\varphi(k+1)\varphi^\top(k+1),\label{eq:RLSsc}
\end{align}
respectively.
\end{remark}
\begin{remark}
One can avoid the computation of a batch LSE \eqref{eq:batchls} completely, by initialization of the recursive LSE \eqref{eq:RLSt} from ``scratch'' with $\hat{\theta}(-1)=0$ and $\Sigma(-1)=c I$,  with $c\in\mathbb{R}_{\geq 0}$ \cite{kay1993}.
\end{remark}

\subsection{Problem formulation}
\begin{figure}
\centering
{\resizebox{2in}{!}{
\begin{tikzpicture}[auto, node distance=3em,>=latex']

\node [block, fill = white, drop shadow](S1) {$\mathcal{I}_1$};
\node [block,below=of S1, fill = white, drop shadow](S2) {$\mathcal{I}_2$};
\node [block,below=of S2, fill = white, drop shadow](S3) {$\mathcal{I}_m$};

\node [hblock,right=of S2, fill = white, drop shadow](B) {$\mathcal{B}$};


\node[input,right=of S1.30] (o1){};
\node[input,right=of S1.330] (s1){};

\node[input,right=of S2.30] (o2){};
\node[input,right=of S2.330] (s2){};

\node[input,right=of S3.30] (o3){};
\node[input,right=of S3.330] (s3){};

\node[input,left=of S1.180] (u1){};
\node[input,left=of S2.180] (u2){};
\node[input,left=of S3.180] (u3){};

\node[input,right=of B] (y){};
\draw[->] (u1) -- node[near start]{$u_1$} (S1.180);
\draw[->] (u2) -- node[near start]{$u_2$} (S2.180);
\draw[->] (u3) -- node[near start]{$u_m$} (S3.180);

\draw [->] (S1.30) -- node{$ $} (o1);
\draw [->] (s1) -- node{$ $} (S1.330);

\draw [->] (S2.30) -- node{$ $} (o2);
\draw [->] (s2) -- node{$ $} (S2.330);

\draw [->] (S3.30) -- node{$ $} (o3);
\draw [->] (s3) -- node{$ $} (S3.330);

\draw [<-] (B) -- node{$y$} (y);

\node at ($(S2)!.45!(S3)$) {\vdots};
%
%
%
\end{tikzpicture}
}}
\caption{Distributed identification scheme with identification modules $\mathcal{I}_i$, $i\in\mathbb{N}_{[1:m]}$ and communication module $\mathcal{B}$.}
\label{fig:distrscheme}
\end{figure}
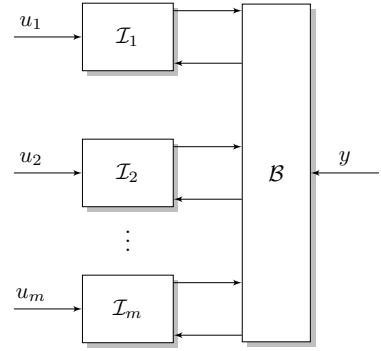

Given the prediction error identification problem for the MISO system described in Section \ref{sec:prederr}, central collection of $m$ input signals $u_i$ and one output signal $y$ is required\footnote{We remark that if all $u_i$'s are uncorrelated, then SISO identification (without modelling other subsystems) provides consistent estimates. This will lead to increased variance, however.} for the central computation of $\hat{\theta}$, using either the LSE \eqref{eq:LSE} or the recursive LSE \eqref{eq:RLSt}. From a distributed point of view, however, local module parameter estimators $\hat{\theta}_i$ for $\theta_i$, may be preferred, due to computational or communication constraints. We will refer to the concept of distributed identification, as the local parameter estimation for $G_i$ via a local identification module, with intercommunication between local identification modules. The distributed identification concept is illustrated in Figure \ref{fig:distrscheme}: Each subsystem $G_i$, $i\in\mathbb{N}_{[1:m]}$ is coupled with an identification module $\mathcal{I}_i$, which measures input $u_i$ and is connected to some module $\mathcal{B}$, which measures the (shared) output $y$. Module $\mathcal{B}$ describes the relation between sent and received signals of all modules $\mathcal{I}_i$. Given this distribution, two problems arise, related to the local identification and communication. Firstly, is there an $\mathcal{I}_i$ that arrives at unbiased estimates of the true parameter $\theta_i^0$? Consequently, if the answer is affirmative, what signals have to be shared between the identification modules $\mathcal{I}_i$, $i\in\mathbb{N}_{[1:m]}$, i.e., what should $\mathcal{B}$ describe?

Since the data matrix $\Phi$ is in general non-sparse, the identification problem $\min_\theta J_{\text{LS}}(\theta)$ is in general non-separable. Therefore, it is not clear how the LSE \eqref{eq:LSE} can be adopted in a distributed identification scheme. The recursive LSE, however, can be advantageous for the distribution of the parameter estimation. Indeed, one can exploit structures for the parameter covariance matrix $\Sigma(k)$, such as diagonal or block-diagonal structures, in order to ``separate'' the estimation problem w.r.t.  $\theta_i$, $i\in\mathbb{N}_{[1:m]}$. Finally, asymptotic unbiasedness of the developed distributed identification procedure should be assessed, i.e., we need to verify whether $\lim_{k\rightarrow \infty} \operatorname{E} \hat{\theta}_i(k)\rightarrow\theta_i^0$, where $\hat{\theta}_i(k)$ denotes the proposed estimator for $\theta_i^0$ based on $k+1$ data samples.

\section{Distributed estimation algorithm} \label{sec:distrp}
Inspired by the recursive LSE \eqref{eq:RLSt}, we develop a distributed recursive estimator: for each $i\in\mathbb{N}_{[1:m]}$, let the local parameter estimator $\hat{\theta}_i:\mathbb{N}\rightarrow\mathbb{R}^{n_i}$ be defined recursively by
\begin{align}
&\hat{\theta}_i(k+1)=\hat{\theta}_i(k) \label{eq:RLSl}\\
&+\alpha_i(k)\Sigma_i(k)\varphi_i(k+1)(y(k+1)-\sum_{j=1}^m\varphi_j(k+1)\hat{\theta}_j(k)),\nonumber
\end{align}
with $\alpha_i:\mathbb{N}\rightarrow\mathbb{R}$ and $\Sigma_i:\mathbb{N}\rightarrow\mathbb{R}^{n_i\times n_i}$. Comparing the local estimator update equation \eqref{eq:RLSl} with the recursive LSE \eqref{eq:RLSt}, the matrix $\Sigma_i$ has the interpretation of a local covariance matrix of $\hat{\theta}_i$. Let $\Sigma_i:\mathbb{N}\rightarrow\mathbb{R}^{n_i\times n_i}$ be defined recursively by
\begin{align} \label{eq:RLSIs}
\Sigma_i^{-1}(k+1)=\Sigma_i^{-1}(k)+\frac{1}{\gamma_i^2(k)}\varphi_i(k+1)\varphi_i^\top(k+1),
\end{align}
with $\gamma_i:\mathbb{N}\rightarrow\mathbb{R}$. The scalars $\alpha_i(k)$ and $\gamma_i(k)$ are related to sufficient conditions for consistency of estimator \eqref{eq:RLSl}, which will be provided in Section 4.

Consider the stacked vector $\hat{\theta}_B(k):=\operatorname{col}(\hat{\theta}_1(k),\dots,\hat{\theta}_m(k))$. Let $A_B(k):=\operatorname{diag}(\alpha_1(k)I_{n_1},\dots ,\alpha_m(k)I_{n_m})$ and let $\Gamma_B(k):=\operatorname{diag}(\gamma_1(k)I_{n_1},\dots ,\gamma_m(k)I_{n_m})$. Define accordingly
\begin{align*}
\Sigma_B(k)&:=\operatorname{diag}(\Sigma_1(k),\dots,\Sigma_m(k)),\\
\varphi_B(k)&:=\operatorname{diag}(\varphi_1(k)\varphi_1^\top(k),\dots,\varphi_m(k))\varphi_m^\top(k).
\end{align*}
For the estimator update we can then write
\begin{align*}
\hat{\theta}_B(k+1)=\hat{\theta}_B(k)+A_B(k)\Sigma_B(k)\varphi(k+1)\varepsilon(k+1,\hat{\theta}_B(k))
\end{align*}
with
\begin{align*}
\Sigma_B^{-1}(k+1)=\Sigma_B^{-1}(k)+\Gamma_B^{-2}(k)\varphi_B(k+1).
\end{align*}
The latter equations seem to resemble \eqref{eq:RLSt} and \eqref{eq:RLSsc}, which describe the recursive LSE. Note, however, that the matrix $\Sigma_B$ is block diagonal, while the covariance matrix $\Sigma$ is dense, in general. 

Now, let identification module $\mathcal{I}_i$ be described by \eqref{eq:RLSl} and \eqref{eq:RLSIs} so that
\begin{align*}
\mathcal{I}_i\hspace{-.25em}:\hspace{-.35em}\left\{\hspace{-.35em}\begin{array}{l}
\hat{\theta}_i(k+1)=\hat{\theta}_i(k)+\alpha_i(k)\Sigma_i(k)\varphi(k+1)\varepsilon(k+1,\hat{\theta}_B(k))\\
\Sigma_i^{-1}(k+1)=\Sigma_i^{-1}(k)+\frac{1}{\gamma_i^2(k)}\varphi_i(k+1)\varphi_i^\top(k+1).
\end{array}\right.
\end{align*}
Writing the distributed estimator \eqref{eq:RLSl} as
\begin{align*}
&\hat{\theta}_i(k+1)=\hat{\theta}_i(k)+\alpha_i(k)\Sigma_i(k)\varphi_i(k+1)\\
&\cdot(y(k+1)-\varphi_i^\top(k+1)\hat{\theta}_i(k)-\hspace{-.8em}\sum_{j\in\mathbb{N}_{[1:m]}\setminus\{i\}}\varphi^\top_j(k+1)\hat{\theta}_j(k)),
\end{align*}
it becomes apparent what information exchange is required between identification modules, assuming that each module $\mathcal{I}_i$ can measure input $u_i(t)$ and receive output $y(t)$. The local recursive estimator $\hat{\theta}_i(k+1)$, $i\in\mathbb{N}_{[1:m]}$, depends on an ``autonomous'' part plus a contribution from other identification modules $\mathcal{I}_j$, $j\in\mathbb{N}_{[1:m]}\setminus\{i\}$. The inputs from other subsystems and parameter vectors $\hat{\theta}_j$ are not required to be known. Indeed, only the scalar products $\varphi^\top_j(k+1)\hat{\theta}_j(k)\in\mathbb{R}$ need to be known, for all $j\in\mathbb{N}_{[1:m]}\setminus\{i\}$, which we will refer to as the local predictions. The appropriate communication can be achieved if, for example, at every time step, each $\mathcal{I}_i$ sends the local prediction $\varphi^\top_i(k+1)\hat{\theta}_i(k)\in\mathbb{R}$ to all other identification modules $\mathcal{I}_j$, $j\in\mathbb{N}_{[1:m]}\setminus\{i\}$ with a corresponding definition for $\mathcal{B}$. The latter corresponds to an all-to-all communication, however, and can be inefficient for large $m$. One can instead consider $\mathcal{B}$ to be described by the static relation
\begin{align*}
\mathcal{B}:\quad \varepsilon(k+1,\hat{\theta}_B(k))=y(k+1)-\sum_{i=1}^m\varphi_i^\top (k+1)\hat{\theta}_i(k)
\end{align*}
and consider the following distributed identification procedure to improve efficiency in the communication:

For all $i\in\mathbb{N}_{[1:m]}$, initialize $\mathcal{I}_i$ at $k=0$ with $\hat{\theta}_i(0)\in\mathbb{R}^{n_i}$ and $0\prec\Sigma_i(0)\in\mathbb{R}^{n_i\times n_i}$. For each time $k\in\mathbb{N}$ perform
\begin{enumerate}
\item[(i)] For each $i\in\mathbb{N}_{[1:m]}$, $\mathcal{I}_i$ measures $u_i(k+1)$ and sends the local prediction $\varphi_i^\top(k+1)\hat{\theta}_i(k)\in\mathbb{R}$ to $\mathcal{B}$.
\item[(ii)] $\mathcal{B}$ measures $y(k+1)$ and returns the prediction error $\varepsilon(k+1,\hat{\theta}_B(k))$ to $\mathcal{I}_i$, $i\in\mathbb{N}_{[1:m]}$.
\item[(iii)] For each $i\in\mathbb{N}_{[1:m]}$, $\mathcal{I}_i$ computes $\hat{\theta}_i(k+1)$ and $\Sigma_i(k+1)$ by \eqref{eq:RLSl} and \eqref{eq:RLSIs}, respectively.
\end{enumerate}

\begin{remark}
The distributed identification procedure can be viewed as a central fusion, distributed computation scheme: local estimations are obtained by modules $\mathcal{I}_i$, $i\in\mathbb{N}_{[1:m]}$, which all connect to $\mathcal{B}$. This scheme reflects the interconnection of the MISO system in Figure \ref{fig:m7scheme}, where all $G_i$, $i\in\mathbb{N}_{[1:m]}$, connect to a single summation point.
\end{remark}

%
%

\section{Convergence analysis} \label{sec:con}
Now that the central and distributed estimators are updated according to \eqref{eq:RLSt} and \eqref{eq:RLSl}, respectively, let us analyze the asymptotic properties of the estimators. In this section, we will first assume perfect measurements of the system output $y(t)$, i.e., the noise signal $v(t)=0$ for $t\in\mathbb{N}$, and that system \eqref{eq:sysmiso} is in the FIR model set, i.e., the output of the real system $y(t)$ can be described by $y(t)=\varphi^\top(t)\theta^0$. We will analyze the desired convergence $\hat{\theta}\rightarrow\theta^0$ via Lyapunov's second method, as was done in the analysis of gradient algorithms for deterministic parameter estimation in \cite{udinktencate78} and \cite{mendel73}. 

\subsection{Central recursive LSE} \label{sec:clse}
We will briefly pay attention to a  convergence result for the recursive LSE, to show the analogy with the convergence result for the distributed recursive estimator in Section \ref{sec:distcon}.

Consider the estimator error $\tilde{\theta}(k):=\hat{\theta}(k)-\theta^0\in\mathbb{R}^n$. In the absence of noise ($v(t)=0$), it follows from \eqref{eq:RLSt} that the recursive LSE error dynamics are described by
\begin{align}
&\tilde{\theta}(k+1)=\tilde{\theta}(k)-\alpha(k)\Sigma(k)\varphi(k+1)\varphi^\top(k+1)\tilde{\theta}(k),\label{eq:RLSte}\\
&\Sigma^{-1}(k+1)=\Sigma^{-1}(k)+\frac{1}{\sigma^2}\varphi(k+1)\varphi^\top(k+1).\label{eq:RLSse}
\end{align}
Observe that the origin is clearly an equilibrium of difference equation \eqref{eq:RLSte}.

\subsubsection{Convergence}
The following result demonstrates that the estimation error converges to zero for the recursive LSE in the deterministic case, i.e., when the noise $v(t)=0$ for all $t\in\mathbb{N}$. A similar result was proven in \cite[Appendix~B]{udinktencate78}, for a least-squares like gradient algorithm.
\begin{proposition} \label{prop:c}
Let $W_C:\mathbb{R}^n\times\mathbb{N}\rightarrow\mathbb{R}$ be defined by
\begin{align*}
W_C(\xi,\tau):=\xi^\top\Sigma^{-1}(\tau)\xi
\end{align*}
and let $\Sigma(k)$ satisfy \eqref{eq:RLSse}, $\Sigma(0)\succ0$. Assume that $\tilde{\theta}(k)$ and $\varphi(k+1)$ are not orthogonal for all $k\in\mathbb{N}$. Then $W_C:\mathbb{R}^n\times\mathbb{N}\rightarrow\mathbb{R}$ is a Lyapunov function for \eqref{eq:RLSte}.
\begin{pf}
See Appendix C.
\end{pf}
\end{proposition}
\begin{remark}
When $\tilde{\theta}(k)$ and $\varphi(k+1)$ are orthogonal, the error system \eqref{eq:RLSte} is stable, but not guaranteed to be asymptotically stable, and convergence cannot be concluded. Orthogonality can, however, always be avoided by utilizing input signals with sufficient independent frequencies \cite{mendel73}.
\end{remark}
\subsection{Distributed recursive estimator} \label{sec:distcon}
\subsubsection{Estimator error dynamics}
Consider the distributed recursive estimator \eqref{eq:RLSl}. When no noise is present in the measured output $y(t)$ ($v(t)=0$), the distributed estimator update \eqref{eq:RLSl} can be written as
\begin{align*}
\hat{\theta}_i(k+1)&=\hat{\theta}_i(k)+\alpha_i(k)\Sigma_i(k)\varphi_i(k+1)\\
&\quad\cdot\left(\sum_{j=1}^m\varphi_j^\top(k+1)\theta^0_j-\sum_{j=1}^m\varphi^\top(k+1)\hat{\theta}_j(k)\right)\\
\end{align*}
where we used $y(k)=\sum_{j=1}^m \varphi_j^\top(k)\theta^0_j$. Now, define the error vector $\tilde{\theta}_B(k):=\hat{\theta}_B(k)-\theta^0\in\mathbb{R}^n$. We then have
\begin{align*}
\tilde{\theta}_B(k+1)&=\hat{\theta}_B(k+1)-\theta^0\\
&=\hat{\theta}_B(k)-\theta^0+A_B(k)\Sigma_B(k)\varphi(k+1)\\
&\quad \cdot \sum_{j=1}^m\varphi_j^\top(k+1)(\theta_j-\hat{\theta}_j(k))\\
&=\tilde{\theta}_B(k)-A_B(k)\Sigma_B(k)\varphi(k+1)\varphi^\top(k+1)\tilde{\theta}_B(k)\\
&=F(k)\tilde{\theta}_B(k)
\end{align*}
with $F(k):=I_n-A_B(k)\Sigma_B(k)\varphi(k+1)\varphi^\top(k+1)$.

Recalling the difference equation for the gain matrix $\Sigma_B(k)$, we conclude that the error behavior of the distributed recursive estimator \eqref{eq:RLSl} is described by
\begin{align}
\tilde{\theta}_B(k+1)&=F(k)\tilde{\theta}_B(k),\label{eq:dRLSte}\\
\Sigma_B^{-1}(k+1)&=\Sigma_B^{-1}(k)+\Gamma_B^{-2}(k)\varphi_B(k+1),\label{eq:dRLSse}
\end{align}
where $\Gamma_B^{-2}(k)=\operatorname{diag}(\gamma_1^{-2}(k)I_{n_1},\dots,\gamma_m^{-2}(k)I_{n_m})$.

\subsubsection{Convergence}
The following result proves the existence of the scalar functions $\gamma_i(k)$ for each estimator, such that the distributed estimation error vector converges to zero in the deterministic case.
\begin{theorem}
\label{thm:dlf}
Let $W_B:\mathbb{R}^n\times \mathbb{N}\rightarrow\mathbb{R}$ be defined by
\begin{align*}
W_B(\xi,\tau):=\xi^\top \Sigma_B^{-1}(\tau)\xi
\end{align*}
and let $\Sigma_B(\cdot)$ satisfy \eqref{eq:dRLSse}, $\Sigma_B(0)\succ0$. For all $i\in\mathbb{N}_{[1:m]}$, let $\alpha_i=\alpha_B$, with $\alpha_B(k):=(\sigma^2+\sum_{j=1}^m \varphi_j^\top (k+1)\Sigma_j(k)\varphi_j(k+1))^{-1}$. Assume that $\tilde{\theta}_B(k)$ and $\varphi(k+1)$ are not orthogonal for all $k\in\mathbb{N}$. Then there exist $\gamma_i$, $i\in\mathbb{N}_{[1:m]}$, such that $W_B:\mathbb{R}^n\times\mathbb{N}\rightarrow\mathbb{R}$ is a Lyapunov function for \eqref{eq:dRLSte}.
\begin{pf}
We will first prove that there exists $k_1\in\mathcal{K}_{\infty}$ s.t. $W_B(\xi,\tau)\geq k_1(\|\xi\|)$ for all $(\xi,\tau)\in\mathbb{R}^n\times \mathbb{N}$, by induction. Let $\xi\in\mathbb{R}^n$ be arbitrary and let $k_1(r):=\lambda_{\min} (\Sigma_B^{-1}(0))r^2$. We claim that $W_B(\xi,\tau)\geq k_1(\|\xi\|)$ for all $\tau\in\mathbb{N}$. For the base case $\tau=0$ the statement is true, since we have $W_B(\xi,0)\geq \lambda_{\min} (\Sigma_B^{-1}(0))\|\xi\|^2=k_1(\|\xi\|)$. Now, let $W_B(\xi,k)\geq k_1(\|\xi\|)$ be true for some $k\in\mathbb{N}$. Then 
\begin{align*}
W_B(\xi,k+1)&=\xi^\top\Sigma_B^{-1}(k+1)\xi\\
&=\xi^\top\Sigma_B^{-1}(k)\xi+\xi^\top\underbrace{\Gamma_B^{-2}(k)\varphi_B(k+1)}_{\geq 0}\xi\\
& \geq \xi^\top\Sigma_B^{-1}(k)\xi\geq k_1(\|\xi\|),
\end{align*}
thus the statement is also true for $k+1$. We conclude that $W_B(\xi,\tau)\geq k_1(\|\xi\|)$ for all $(\xi,\tau)\in\mathbb{R}^n\times \mathbb{N}$. For the upperbound, let  $k_2(r,k):=\lambda_{\max} (\Sigma_B^{-1}(k))r^2$. Then $W_B(\xi,\tau)\leq \lambda_{\max}(\Sigma_B^{-1}(\tau))\|\xi\|^2=k_2(\|\xi\|,\tau)$ for all $\xi\in\mathbb{R}^n$.

Let us now analyze the one-step-difference $\Delta W_B(k):=W_B(\tilde{\theta}_B(k+1),k+1)-W_B(\tilde{\theta}_B(k),k)$. Using the distributed estimator error dynamics \eqref{eq:dRLSte}, we find
\begin{align*}
\Delta W_B(k)&=\tilde{\theta}_B^\top(k+1)\Sigma_B^{-1}(k+1)\tilde{\theta}_B(k+1)\\
&\quad-\tilde{\theta}_B^\top(k)\Sigma_B^{-1}(k)\tilde{\theta}_B(k)\\
&=\tilde{\theta}_B^\top(k+1)\left(\Sigma_B^{-1}(k+1)-\Sigma_B^{-1}(k)\right)\tilde{\theta}_B(k+1)\\
+&\tilde{\theta}_B^\top(k+1)\Sigma_B^{-1}(k)\tilde{\theta}_B(k+1)-\tilde{\theta}_B^\top(k)\Sigma_B^{-1}(k)\tilde{\theta}_B(k)\\
&=\overline{\Delta W}_B(k)\\
&\quad +\tilde{\theta}_B^\top(k+1)\left(\Sigma_B^{-1}(k+1)-\Sigma_B^{-1}(k)\right)\tilde{\theta}_B(k+1),
\end{align*}
where
\begin{align*}
\overline{\Delta W}_B&:=\tilde{\theta}_B^\top(k+1)\Sigma_B^{-1}(k)\tilde{\theta}_B(k+1)-\tilde{\theta}_B^\top(k)\Sigma_B^{-1}(k)\tilde{\theta}_B(k)\\
&={\theta}_B^\top\Sigma_B^{-1}\tilde{\theta}_B-2\tilde{\theta}_B^\top\varphi\varphi^\top\Sigma_BA_B\Sigma_B^{-1}\tilde{\theta}_B\\
&\quad+\tilde{\theta}_B^\top\varphi\varphi^\top\Sigma_BA_B\Sigma_B^{-1}A_B\Sigma_B\varphi\varphi^\top\tilde{\theta}_B\\
&\quad-{\theta}_B^\top\Sigma_B^{-1}\tilde{\theta}_B\\
&=\tilde{\theta}_B^\top\varphi\varphi^\top\Sigma_BA_B\Sigma_B^{-1}A_B\Sigma_B\varphi\varphi^\top\tilde{\theta}_B\\
&\quad-2\tilde{\theta}_B^\top\varphi\varphi^\top\Sigma_BA_B\Sigma_B^{-1}\tilde{\theta}_B.
\end{align*}
Now, since $\alpha_i=\alpha_B$ for all $i\in\mathbb{N}_{[1:m]}$, we have that $\overline{\Delta W}_B$ simplifies to
\begin{align*}
\overline{\Delta W}_B&=\alpha_B^2\tilde{\theta}_B^\top\varphi\varphi^\top\Sigma_B\varphi\varphi^\top\tilde{\theta}_B-2\alpha_B\tilde{\theta}_B^\top\varphi\varphi^\top\tilde{\theta}_B\\
&=-\alpha_B(\tilde{\theta}_B^\top\varphi)^2\left(2-\alpha_B\varphi^\top\Sigma_B\varphi\right),
\end{align*}
so that $\overline{\Delta W}_B$ is negative when
\begin{align*}
0<\alpha_B<\frac{2}{\varphi^\top\Sigma_B\varphi}.
\end{align*}
Since $\alpha_B=(\sigma^2+\sum_{j=1}^m \varphi_j^\top \Sigma_j\varphi_j)^{-1}$, the latter condition is satisfied, such that $\overline{\Delta W}_B<0$.

By equation \eqref{eq:dRLSse}, the one-step-difference is equal to
\begin{align*}
\Delta W_B(k)&=\overline{\Delta W}_B+\tilde{\theta}_B^\top(k+1)\Gamma_B^{-2}(k)\varphi_B(k+1)\tilde{\theta}_B(k+1)\\
&=\overline{\Delta W}_B+\sum_{i=1}^m\frac{1}{\gamma_i^2(k)}\tilde{\theta}^\top_i(k+1)\varphi_i\varphi_i^\top\tilde{\theta}_i(k+1)\\
&\leq \overline{\Delta W}_B+\sum_{i=1}^m\frac{1}{\gamma_i^2(k)}\sum_{j=1}^m(\tilde{\theta}^\top_j(k+1)\varphi_j)^2,
\end{align*}
where we used the Cauchy-Schwarz inequality. The decrease condition $\Delta W_B(k)<0$ is therefore satisfied whenever $\gamma_i(k)$, $i\in\mathbb{N}_{[1:m]}$, are so large that
\begin{align*}
\sum_{i=1}^m\frac{1}{\gamma_i^2(k)}<\frac{\overline{|\Delta W}_B|}{\tilde{\theta}_B^\top(k)F^\top(k)\varphi_BF(k)\tilde{\theta}_B(k)}
\end{align*}
for all $k\in\mathbb{N}$, which is equivalent to the existence of $k_3:\mathbb{R}_{\geq 0}\rightarrow\mathbb{R}_{\geq 0}$ such that \eqref{cond:2b} holds \cite{malisoff2009}. This concludes the proof.
\end{pf}
\end{theorem}
\begin{remark}
The difference in the stability analysis of the distributed estimator w.r.t. the recursive LSE is induced by \eqref{eq:dRLSse}. The block-diagonality of $\varphi_B$ requires conditions on $\gamma_i$ for stability, whereas a multiplication of $\varphi\varphi^\top$ with $\sigma^{-2}$ in \eqref{eq:RLSse} suffices for stability of the recursive LSE.
\end{remark}
\begin{remark}
The proof of Theorem \ref{thm:dlf} gives exact conditions on the scalar functions $\gamma_i(k)$. When $\gamma_i(k)=\gamma_i$ is chosen to be a constant, it suffices to assume that $\gamma_i\in\mathbb{R}$ is large enough, such that $\Delta W_B(k)<0$. 
\end{remark}

In the presence of noise, the error dynamics for the distributed estimator are described by
\begin{align*}
\tilde{\theta}_B(k+1)&=F(k)\tilde{\theta}_B(k)+G(k)v(k+1),
\end{align*}
with $G(k):=A_B(k)\Sigma_B(k)\varphi(k+1)$. The following result provides sufficient conditions for asymptotic unbiasedness of the distributed estimator $\hat{\theta}_B$.

\begin{proposition}
Let $\prod_{\tau=t+1}^k F(\tau)G(t)$ and $v(t)$ be statistically independent for all $t\in\mathbb{N}$. If there exists a Lyapunov function for \eqref{eq:dRLSte}, then $\lim_{k\rightarrow\infty} \operatorname{E} \hat{\theta}_B(k)=\theta^0$.
\begin{pf}
We refer the reader to the  proof of \cite[Theorem~2-5]{mendel73}.
\end{pf}
\end{proposition}

\section{Numerical example}
Consider the data generating system \eqref{eq:sysmiso} with $m=20$ subsystems, so that $y(t)=\sum_{i=1}^{20} B_i(q)u_i(t)+v(t)$,  with $B_i(q)=b_0^i+b_1^iq^{-1}+\cdots + b_{n_i-1}^iq^{-n_i+1}$ and $v(t)$ zero-mean white Gaussian noise with standard deviation $\sigma=0.1$. For this illustrative example, the subsystems $G_i(q)$ of the data generating system are constructed in a random fashion as follows: each subsystem has $n_i\in\mathbb{N}$ unknown parameters, which is an integer drawn from a discrete uniform distribution $\mathcal{U}\{1,10\}$ using the \textsc{Matlab} function \texttt{randi}, i.e., subsystem $G_i$ has $n_i\in\mathbb{N}_{[1:10]}$ unknown parameters. The constant parameters $b_j^i\in\mathbb{R}$, $j\in\mathbb{N}_{[0:n_i-1]}$, $i\in\mathbb{N}_{[1:20]}$, are drawn from a normal distribution $\mathcal{N}(0,1)$ in \textsc{Matlab} using \texttt{randn}. The total number of to-be-estimated parameters is $n=\sum_{i=1}^{20}n_i=102$.

We apply the distributed recursive estimation procedure from Section \ref{sec:distrp}. The local estimators $\hat{\theta}_i:\mathbb{N}\rightarrow\mathbb{R}^n_i$ are described by \eqref{eq:RLSl} with $\alpha_i=\alpha_B$, $i\in\mathbb{N}_{[1:20]}$, as defined in Theorem \ref{thm:dlf}. The matrices $\Sigma_i:\mathbb{N}\rightarrow\mathbb{R}^{n_i\times n_i}$ are described by \eqref{eq:RLSIs}, with $\gamma_i(k)=\gamma=100$. For comparison, we apply a corresponding central recursive estimator, i.e., the recursive LSE \eqref{eq:RLSt} with the update for the matrix $\Sigma:\mathbb{N}\rightarrow\mathbb{R}^{n\times n}$ described by $\Sigma^{-1}(k+1)=\Sigma^{-1}(k)+\gamma^{-2}\varphi(k+1)\varphi^\top(k+1)$ instead of \eqref{eq:RLSsc}. 

\begin{figure}[!t]
\centering
\includegraphics[width=3.5in]{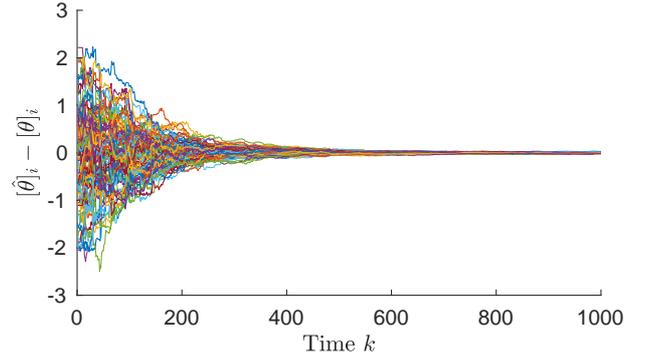}
\caption{Evolution of the estimation error for all parameters $[\hat{\theta}]_j-[\theta^0]_j$, $j\in\mathbb{N}_{[1:102]}$, for the central identification of a MISO system with $m=20$ subsystems.}
\label{fig:ec}
\end{figure}

\begin{figure}[!t]
\centering
\includegraphics[width=3.5in]{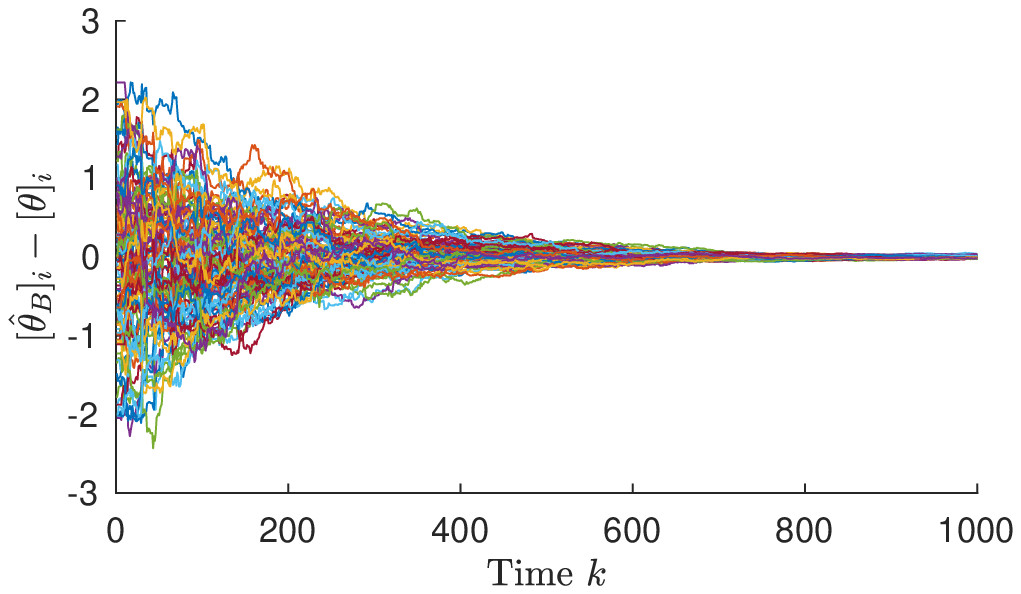}
\caption{Evolution of the estimation error for all parameters $[\hat{\theta}_B]_j-[\theta^0]_j$, $j\in\mathbb{N}_{[1:102]}$, for the distributed identification of a MISO system with $m=20$ subsystems.}
\label{fig:ed}
\end{figure}

\begin{figure}[!t]
\centering
\includegraphics[width=3.5in]{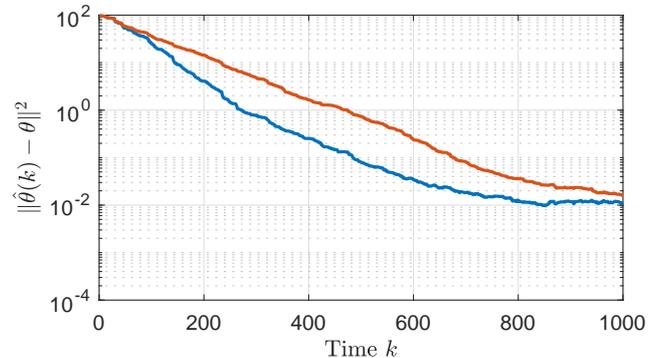}
\caption{Estimation error for the central estimator $\|\tilde{\theta}(k)\|^2$ (blue) and distributed estimator $\|\tilde{\theta}_B(k)\|^2$  (red) for $\gamma=100$.}
\label{fig:en}
\end{figure}

Figure \ref{fig:ec} and \ref{fig:ed} show the evolution of the estimation error over time for the central and distributed estimator, respectively, initialized in $\hat{\theta}(1)=\hat{\theta}_B(1)=0$ and $\Sigma(1)=\Sigma_B(1)=100I_{102}$. The overall estimation errors $\|\hat{\theta}(k)-\theta^0\|^2$ and $\|\hat{\theta}_B(k)-\theta^0\|^2$ are shown in Figure \ref{fig:en} in blue and red, respectively. We observe a lower decrease rate for the estimation errors in the distributed identification scheme w.r.t. the central scheme, in general, while convergence is observed for both schemes.



\section{Conclusions}
We have stated a recursive estimation algorithm for the distributed identification of MISO FIR systems, derived from a central recursive least squares estimator. The distributed identification scheme consists of local identification modules, which estimate a subvector of the total parameter vector. Via Lyapunov's second method, we have obtained sufficient conditions for asymptotic convergence of the estimators to the true parameters in the absence of noise, which leads to asymptotic unbiasedness in the presence of a white noise signal at the system's output.


\appendix
\section{Proof of  Theorem \ref{thm:stab}}
\begin{pf}
The proof follows the same line of reasoning as the proof for the \emph{continuous-time} version of the theorem \cite[Section~5.3.1, Theorem~1]{VidyasagarNonlinear}. We give the proof for completeness.

Let $\varepsilon>0$ and $t_0\in\mathbb{N}$ be given. We will show that there exists $\delta(\varepsilon,t_0)>0$ so that
\begin{align*}
\|x_0\|<\delta(\varepsilon,t_0)\ \Rightarrow\ \|s(t,t_0,x_0)\|<\varepsilon\quad \forall t\geq t_0.
\end{align*}
Take $\delta=\delta(\varepsilon,t_0)>0$ so that
\begin{align*}
k_2(\delta,t_0)<k_1(\varepsilon).
\end{align*}
Such a $\delta>0$ always exists, since $k_1(\varepsilon)>0$ and $k_2(\delta,t_0)\rightarrow 0$ as $\delta\rightarrow 0$. Now, let $\|x_0\|<\delta$. Then
\begin{align*}
W(x_0,t_0)\leq k_2(\delta,t_0)<k_1(\varepsilon).
\end{align*}
From \eqref{cond:1b}, it follows that for all $t\geq t_0$ we have
\begin{align*}
W(s(t,t_0,x_0),t)\leq W(x_0,t_0).
\end{align*}
Since  $W(s(t,t_0,x_0),t)\geq k_1(\|s(t,t_0,x_0)\|)$ by \eqref{cond:1a}, we have
\begin{align*}
k_1(\|s(t,t,_0,x_0\|)&\leq W(s(t,t_0,x_0),t)\\
&\leq W(x_0,t_0)\leq k_2(\delta,t_0)<k_1(\varepsilon),
\end{align*}
which implies
\begin{align*}
\|s(t,t_0,x_0)\|<\varepsilon,\quad \forall t\geq t_0.
\end{align*}
Therefore, the origin equilibrium of \eqref{eq:fdt} is stable, which concludes the proof.
\end{pf}

\section{Proof of Theorem \ref{thm:lyap}}
\begin{pf}
Let $t_0\in\mathbb{N}$ and $x_0\in\mathbb{R}^n$. Since \eqref{cond:2a} and \eqref{cond:2b} imply conditions \eqref{cond:1a} and \eqref{cond:1b}, the origin is stable by Theorem \ref{thm:stab}. It remains to be proven that the origin is globally attractive, i.e., $\lim_{t\rightarrow\infty} s(t,t_0,x_0)=0$.

Suppose that the origin is not attractive, i.e., 
\begin{align*}
\neg\left[\lim_{t\rightarrow\infty} \|s(t,t_0,x_0)\|=0\right]
\end{align*}
is true ($\neg a$ denotes the negation of assertion $a$). Then there is a sufficiently small positive number $c\in\mathbb{R}$ so that $\|s(t,t_0,x_0)\|\geq c$ for all $t\geq t_0$. Indeed, suppose such a number does not exist if $\neg[\lim_{t\rightarrow\infty} \|s(t,t_0,x_0)\|=0]$. Then for some $t^\star\geq t_0$, we must have $\|s(t^\star,t_0,x_0)\|=0$, which implies $s(t^\star,t_0,x_0)=0$. But then $f(s(t^\star,t_0,x_0),t^\star)=f(0,t^\star)=0$, which implies $s(t,t_0,x_0)=0$ for all $t\geq t^\star$. This contradicts the assertion $\neg[\lim_{t\rightarrow\infty} s(t,t_0,x_0)=0]$, hence there exists such a number $c>0$ so that $\|s(t,t_0,x_0)\|\geq c$ for all $t\geq t_0$.

Since $\|s(t,t_0,x_0)\|\geq c$, we have that $k_1(\|s(t,t_0,x_0)\|)\geq k_1(c)$ and hence 
\begin{align*}
W(s(t,t_0,x_0),t)\geq k_1(c)>0
\end{align*}
for all $t\geq t_0$, by \eqref{cond:2a}. Define a non-decreasing and positive-definite function $\rho:\mathbb{R}_{\geq 0}\rightarrow\mathbb{R}_{\geq 0}$ by
\begin{align*}
\rho(s)=\inf_{z\geq s} k_3(z),
\end{align*}
so that $\rho(s)\leq k_3(s)$ for all $s\in\mathbb{R}_{\geq 0}$. Then $\rho(\|s(t,t_0,x_0)\|)\geq \rho(c)$, since $\rho:\mathbb{R}_{\geq 0}\rightarrow\mathbb{R}_{\geq 0}$ is non-decreasing, so that
\begin{align*}
\Delta W(s(t,t_0,x_0),t)\leq -\rho(c)
\end{align*}
for all $t\geq t_0$, by \eqref{cond:2b}. Therefore, we find
\begin{align*}
W(s(t,t_0,x_0),t)&=W(x_0,t_0)+\sum_{\tau=t_0}^t \Delta W(s(\tau,t_0,x_0),\tau)\\
&\leq W(x_0,t_0)+\sum_{\tau=t_0}^t -\rho(c)\\
&=W(x_0,t_0)-\rho(c)(t-t_0)
\end{align*}
and hence
\begin{align*}
0<k_1(c)\leq W(s(t,t_0,x_0),t)\leq W(x_0,t_0)-\rho(c)(t-t_0).
\end{align*}
For sufficiently large values of $t$, the right-hand side of the latter inequality becomes negative, which cannot be true. Therefore, we conclude that $\lim_{t\rightarrow\infty} s(t,t_0,x_0)=0$, which concludes the proof.
\end{pf}

\section{Proof of Proposition \ref{prop:c}}
\begin{pf}
We will first prove condition \eqref{cond:2a}. Let $\xi\in\mathbb{R}^n$ be arbitrary and let $k_1(r):=\lambda_{\min}(\Sigma^{-1}(0))r^2$. We claim that $W_C(\xi,\tau)\geq k_1(\|\xi\|)$ for all $\tau\in\mathbb{N}$. For the base case $\tau=0$ the statement is true, since we have $W_C(\xi,0)\geq \lambda_{\min}(\Sigma^{-1}(0))\|\xi\|^2=k_1(\|\xi\|)$. Now, let $W_C(\xi,k)\geq k_1(\|\xi\|)$ be true for some $k\in\mathbb{N}$. Then
\begin{align*}
W_C(\xi,k+1)&=\xi^\top \Sigma^{-1}(k+1)\xi\\
&=\xi^\top\Sigma^{-1}(k)\xi+\left(\frac{\xi^\top\varphi(k+1)}{\sigma}\right)^2\geq k_1(\xi),
\end{align*}
thus the statement is also true for $k+1$. We conclude that $W_C(\xi,\tau)\geq k_1(\|\xi\|)$ for all $(\xi,\tau)\in\mathbb{R}^n\times\mathbb{N}$. For the upperbound, let $k_2(\xi)(r,k):=\lambda_{\max}(\Sigma^{-1}(k))r^2$. Then $W_C(\xi,\tau)\leq \lambda_{\max}(\Sigma^{-1}(\tau))\|\xi\|^2=k_2(\|\xi\|,\tau)$ for all $\xi\in\mathbb{R}^n$.

We investigate the one-step difference $\Delta W_C(k):=W_C(\tilde{\theta}(k+1),k+1)-W_C(\tilde{\theta}(k),k)$. Using the estimator error dynamics \eqref{eq:RLSte}, we find that
\begin{align*}
\Delta W_C(k)&=\tilde{\theta}^\top(k+1)\Sigma^{-1}(k+1)\tilde{\theta}(k+1)\\
&\quad-\tilde{\theta}^\top(k)\Sigma^{-1}(k)\tilde{\theta}(k)\\
&=\tilde{\theta}^\top(k)\Sigma^{-1}(k+1)\tilde{\theta}(k)-\tilde{\theta}^\top(k)\Sigma^{-1}(k)\tilde{\theta}(k)\\
&+\alpha^2(k)\tilde{\theta}^\top(k)\varphi(k+1)\varphi^\top(k+1)\Sigma(k)\Sigma^{-1}(k+1)\\
&\quad\cdot\Sigma(k)\varphi(k+1)\varphi^\top(k+1)\tilde{\theta}(k)\\
-2&\tilde{\theta}^\top(k)\Sigma^{-1}(k+1)\alpha(k)\Sigma(k)\varphi(k+1)\varphi^\top(k+1)\tilde{\theta}(k).
\end{align*}

Substituting the covariance matrix update equation \eqref{eq:RLSse} into the latter equation, we determine that
\begin{align*}
\Delta W_C(k)&=(\tilde{\theta}\varphi)^2\left(\frac{1}{\sigma^2}+\alpha^2\varphi^\top\Sigma\varphi+\frac{\alpha^2}{\sigma^2}\left(\varphi^\top\Sigma\varphi\right)^2-2\alpha\right.\\
&\qquad \left.-2\frac{\alpha}{\sigma^2}\varphi^\top\Sigma\varphi\right),
\end{align*} 
where we omitted the time dependence of the variables on the RHS for brevity. Recalling the definition of $\alpha(k)$, we can further rewrite $\Delta W(k)$ as
\begin{align*}
 \Delta W_C(k)=-\frac{\tilde{\theta}^\top\varphi\varphi^\top\tilde{\theta}}{\sigma^2+\varphi^\top\Sigma\varphi}.
\end{align*}
It is now easily seen that
\begin{align*}
\Delta W_C(k)=-\frac{\tilde{\theta}^\top(k)\varphi(k+1)\varphi^\top(k+1)\tilde{\theta}(k)}{\sigma^2+\varphi^\top(k+1)\Sigma(k)\varphi(k+1)}<0
\end{align*}
if $\tilde{\theta}(k)^\top\varphi(k+1)\neq 0$, which implies the existence of $k_3:\mathbb{R}_{\geq 0}\rightarrow\mathbb{R}_{\geq 0}$ so that \eqref{cond:2b} holds \cite{malisoff2009}, which concludes the proof.
\end{pf}

\bibliographystyle{IEEEtran}
\bibliography{./distridentext}
\end{document}